\documentclass[twocolumn,letterpaper]{IEEEAerospaceCLS}  

\usepackage{graphicx}    
\newcommand{\ignore}[1]{}  

\usepackage{cite}
\usepackage{xcolor}

\usepackage{amsmath,amssymb,amsfonts}
\usepackage{algorithmic}
\usepackage{textcomp}
\usepackage{xcolor}
\def\BibTeX{{\rm B\kern-.05em{\sc i\kern-.025em b}\kern-.08em
		T\kern-.1667em\lower.7ex\hbox{E}\kern-.125emX}}
\newtheorem{remark}{Remark}

\usepackage{comment}
\usepackage[capitalize]{cleveref}

\begin{document}
\title{FPGA prototyping of synchronized chaotic map for UAV secure communication}

\author{%
Christian Nwachioma\\ 
Center for Innovation \& Tech. Development in Computing\\
National Polytechnic Institute\\
Mexico City, 07700\\
christian.nwachioma@gmail.com
\and 
Martins Ezuma\\
Department of Electrical Engineering\\
North Carolina State University\\
Raleigh, North Carolina USA\\
mcezuma@ncsu.edu
\and
Olusiji .O. Medaiyese\\
Department of Computer Science and Engineering\\
University of Louisville\\
Louisville, Kentucky USA\\
o0meda01@louisville.edu
\thanks{\footnotesize 978-1-7281-7436-5/21/$\$31.00$ \copyright2021 IEEE}              
}

\maketitle

\thispagestyle{plain}
\pagestyle{plain}

\maketitle

\thispagestyle{plain}
\pagestyle{plain}

\begin{abstract}
We propose a security architecture that uses the principle of chaos for UAV secure communication. A UAV, identified as an aerial base station (ABS), communicates with a ground base station (GBS) over a wireless radio frequency (RF) channel. The communication units of the ABS and GBS have dynamics according to the logistic map. The map is chaotic in the appropriate parameter space.  Its states are non-periodic, broadband, and noise-like in the frequency domain. They are useful for spreading information data during transmission, making it extremely difficult for an eavesdropper to recover the modulated message since state prediction is ultimately impossible. To retrieve it, we propose a variable feedback controller. We prove that it can asymptotically stabilize the error dynamics when the information source is off. During transmission, the controller synchronizes the units such that the error contains signatures of the information signal. Therefore, the information signal is retrievable by a suitable detection mechanism. Security depends on the confidentiality of the map, the variable feedback controller, including its scale factor and bounded feedback gain and the designer’s choice of invertible function for use in the scrambling and descrambling process. Also, the method is less prone to jamming attacks and multipath effects as the broadband spectrum can be used to randomly select RF channels. It uses only a few simple algorithms, including a correlation summation and a detection mechanism. The algorithms collect subsamples of the received signal sequences and averages over each subsample length. The method requires minimal programming efforts and low hardware resource utilization. It is energy-efficient, which is a vital consideration for any UAV security model. Moreover, we realize a prototype of the communication system on field-programmable gate arrays (FPGAs). We presented a digital design of the secure communication system involving the transmission of bitstreams between the ABS and GBS.
\end{abstract}

\tableofcontents

\section{Introduction}
The emergence of Unmanned Aerial Vehicles (UAV) holds promise in activities like delivery service, forestry management, public safety, search/rescue operation, firefighting, and agriculture~\cite{shakhatreh2019unmanned}. Besides, UAVs have become an integral part of modern military arsenals. Security and safety are among the setbacks in widely deploying them~\cite{yaacoub2020security}. In~\cite{ezuma2019micro,ezuma2019detection}, a simple passive radio frequency (RF) sensor detects and captures the radio control signals from the ground base station (GBS) to a UAV and vice-versa.  Dangers abound if an airborne UAV is unethically hacked as it communicates with the GBS. In the worst-case scenario, such a compromised UAV could be used for malicious intentions. Security and privacy concerns have motivated relevant regulatory organizations and researchers to propose safety protocols.  

\begin{figure}[t]
	\centerline{\includegraphics[width=0.8\linewidth]{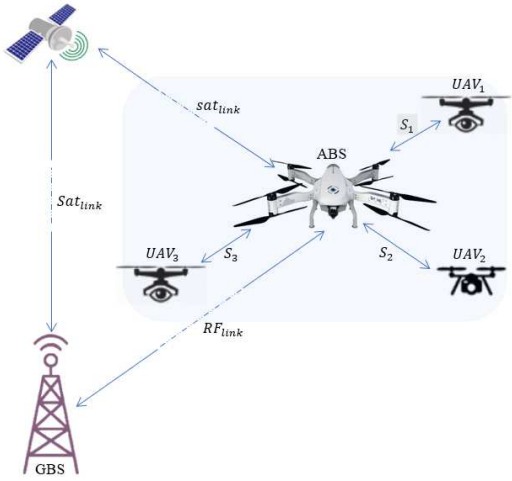}}
	\caption{Transmissions via the $RF_{link}$ and/or $sat_{link}$ are scrambled by states of the logistic map.}
	\label{Fig1} 
\end{figure}

Recently, security standards, including DO-178C and ISO 21384-3:2019, are updated for improved safety of Unmanned Aircraft Systems (UAS)~\cite{IDISOUAV2019}. The standards are based on logic and algorithms put forward by human designers. Hence, their implementation is bug-prone. Besides, it often requires a sizable allocation of microprocessor and RAM and lots of programming efforts. This situation significantly contributes to the energy demands of UAVs. This work intends to offer a perspective that could complement existing UAV security standards and potentially make them more robust in terms of hardware resource utilization and energy-efficiency as inferred above. We propose using the inherent randomness in chaos to provide security for UAS, specifically between a UAV identified as an aerial base station (ABS) in Fig.~\ref{Fig1} and a GBS. In addition to the advantages stated earlier, the proposed security architecture is largely non-algorithmic, cf.~\cite{faraji2020secure}. Thus, code-based hacking is unlikely.

The rest of the paper is organized as follows: Section~\ref{two} provides some background on chaos and control. Section~\ref{three} describes the logistic map and proposes a variable feedback controller for synchronizing the map in the chaotic domain. Section~\ref{five} describes the UAV secure communication design. Section~\ref{six} presents the simulation studies. Section~\ref{seven} presents FPGA prototyping of the proposed method.

\vspace{-5mm}
\section{Background on chaos and control}\label{two}
Deterministic chaos is synthesizable from simple nonlinear dynamical systems. Despite being chaotic, tracking the trajectory is possible using the principle of control theory~\cite{pecora2015synchronization}. This has occasioned applications in secure communication\cite{bendoukha2019secure}, random number synthesis\cite{yu2019survey}, robotics\cite{moysis2020chaotic}, and so on. Although early on chaotic systems were used in the study of population\cite{may1976simple}, weather\cite{lorenz1963deterministic} and economic\cite{holyst1996control} models, they have in recent years, been applied to science and engineering products\cite{Nwachioma2019New,nwachioma2017realization}.

A nonlinear dynamical system is chaotic if it has a positive Lyapunov exponent, dense periodic orbit, and topological transitivity. Hence, infinitesimally close trajectories will become increasingly uncorrelated yet bounded. Thus, making long-term prediction impossible. This behavior is commonly called sensitive dependence on initial conditions~\cite{can2020new}.

A nonlinear continuous-time system necessarily has to be at least $3$-dimensional to exhibit chaos~\cite{sprott2010elegant}. For a discrete-time system (or map) of degree $q$, where $q>0$ is an integer, $1$-dimension will suffice. There are several control schemes in the literature for continuous-time systems despite being bulky. The reason may be due to the relative ease of designing the controllers using some forms of input transformations and Lyapunov theory. Besides, when the system parameter is known, it is easy to realize a Lyapunov rate that is strictly negative. Also, global asymptotic stability follows straightforwardly.

In~\cite{ali1997ssynchronization}, feedback controllers synchronized the logistic map in the presence of noise. However, the controllers were come about by trial and error techniques, making them unreliable. Besides, only a few tabulated numerical data validated synchronization. Also, in~\cite{morgul1998synchronization}, a variable feedback controller was obtained by an analytical method. The effect of the bounded noise was handled in a way that the synchronization error remained within some predefined boundary. In both cases, an advanced notion of stability was unattained. 

Comparing maps to their continuous-time equation counterparts, there are much fewer digital implementations involving them despite their simplicity and promise in applications like communication encryption and random number generation.

Considering the ease of digital implementation of chaotic maps, they have not received the needed attention when it comes to controller design and the corresponding engineering use. The reason may be due to the challenges involved in proving advanced notions of stability. However, for the logistic map, it turns out that this is straightforward.
\vspace{-5mm}
\section{Synchronized logistic map}\label{three}
\subsection{System description}
\label{logisticmap}
The logistic map is a first-order nonlinear difference equation\cite{may1976simple}. The map can be described as
\begin{equation}
x_{n+1} = \mu{x_n}\bigg(1-{x_n\over k}\bigg)\label{drive},
\end{equation}
where $n=0,1,2,...$, $k>0$, is a scale factor, $x_n$ are states of the map and lie in the set $(0,k)$. In the canonical form, $k=1$. The system becomes unstable if $x_n$ goes outside the set. Hence, we may liken the set to the basin of attraction\cite{perez2020luenberger}. The control parameter $\mu$ is usually varied between $0$ and $4$. The map exhibits different interesting behaviors as the parameter is varied. Slowly increasing the parameter, a geometric increase in degrees of periodicity arises. When $\mu\ge3.57$, the map has dense and infinite periodic orbits. That is, periodicity can no longer be seen in the dynamics. From this point, the system is said to be chaotic. 

\begin{figure}[htbp]
	\centerline{\includegraphics[width=0.9\linewidth]{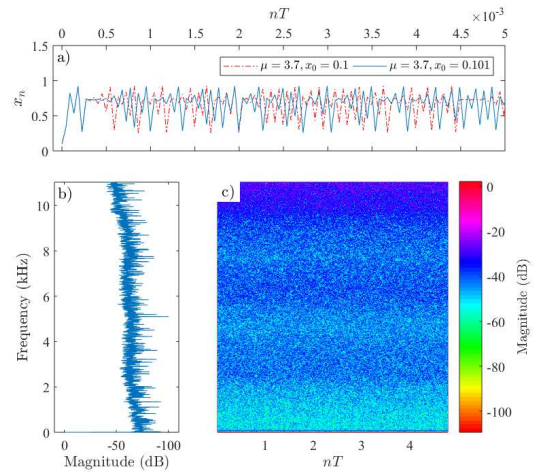}}
	\caption{\textbf{(a)} Time domain signals of the logistic map for $\mu=3.7$, started at slightly different initial conditions as indicated. \textbf{(b)} Amplitude spectrum of the signal shows that there is no frequency component for $\mu=3.7$. \textbf{(c)} Corresponding spectrogram corroborates the previous result.}
	\label{Fig2}
\end{figure}

Choosing the parameter $\mu=3.7$, Fig.~\ref{Fig2}{(a)} exhibits sensitive dependence on the indicated initial conditions. Fig.~\ref{Fig2}{(b)} shows a frequency spectrum derived from the time-series data when $\mu=3.7$ and $x_0=0.1$. The figure shows that no frequency component is present in the spectrum. Fig.~\ref{Fig2}{(c)} is a spectrogram that further corroborates the previous result as no sudden change in magnitude is present. Hence, although the time-series data is from a deterministic system, it has infinite periodic orbits and is noise-like. Thus, the logistic map can be used to scramble a signal during transmission. 
\vspace{-2mm}
\subsection{Synchronization}\label{four}
\label{sync} 
Let us take system \eqref{drive} to be the communication unit of the GBS and we describe the communication unit of the ABS as
\begin{equation}
y_{n+1} = \mu y_n\bigg(1-{y_n\over k}\bigg)+u(e_n,x_n)\label{response}.
\end{equation}
Unlike the states $x_n$ of the GBS communication unit, states of the ABS communication unit $y_n$, does not necessarily have to be confined to $(0,k)$. We want to realize synchronization of the chaotic states of the units over an RF channel so that we can achieve secure information exchange over states of the units. Hence, the synchronization error of the coupled units can be given by
\begin{equation}
e_n:=y_n-x_n\label{syncerror}.
\end{equation}
For any $x_0\neq y_0$ and $x_0+y_0\neq{k}$, our aim is to design and deploy a \textit{variable feedback controller} $u(e_n,x_n)$ such that
\begin{equation}
\lim\limits_{n\rightarrow\infty}|e_n|=0\label{aim}.
\end{equation} 
In a real system, $x_0\neq{y_0}$ is often the case due to sensitive dependence on initial conditions as illustrated in \cref{logisticmap}. For the same reason, $x_0+y_0=k$ is very unlikely. But assuming we can choose the initial conditions such that
\begin{equation}
x_i+y_i=k\label{condic},
\end{equation}
where $i$ is any point in the set $n=0,1,2,..$, then there would be synchronization from the $(i+1)^{th}$ samples since
\begin{equation}
\begin{split}
x_{i+1}=\mu{x_i}\bigg(1-{x_i\over{k}}\bigg)=\mu{x_i}{y_i\over{k}}=\mu{y_i\over{k}}(k-y_i)=y_{k+1},
\end{split}
\end{equation}
in which case there would be no need for a controller. However, in practice, it is more feasible to deploy a controller than to choose initial conditions for which \eqref{condic} is satisfied. Hence, for synchronization of the units in the chaotic regime, we propose the following variable feedback controller:
\begin{equation}
u(e_n,x_n) = \bigg[\mu(e_n+2x_n-k)+\rho{k}\bigg]{e_n\over{k}},\label{controlA}	
\end{equation}
where $\rho$ is the feedback gain. Advancing the maps, the error advances as
\begin{equation}
\begin{split}
e_{n+1} &=y_{n+1}-x_{n+1}\\
&=\mu e_n\bigg[1-{1\over{k}}\bigg(x_n+y_n\bigg)\bigg]+u(e_n,x_n)\label{openerror}.
\end{split}
\end{equation}
Using the control input transformation \eqref{controlA}, we have
\begin{equation}
e_{n+1}=\rho e_{n}\label{linearerrorA}.
\end{equation}
While recursive iteration can be used to solve \eqref{linearerrorA}, we shall instead use the Lyapunov method as it allows us to establish a stronger notion of stability. First, we define a positive definite function:
\begin{equation}
V_n = e_n^2 >0\enspace\mathtt{in}\enspace D-\{0\},
\end{equation}
where $D\subset\mathbb{R}$ is some domain of interest. A subsequent iterate on $V_n$ with respect to the dynamics \eqref{linearerrorA} will be
\begin{equation}
V_{n+1} = \rho^2e_n^2.
\end{equation}
The change between an immediate future state and the current state is
\begin{equation}
V_{n+1}-V_n=-e_n^2(1-\rho^2)\label{dv}.
\end{equation}
Therefore, whenever the feedback gain $|\rho|\le1$, the error system is stable in the sense of Lyapunov. For asymptotic stability, $|\rho|<1$. Moreover, since the Lyapunov function candidate is radially unbounded, i.e.
\begin{equation}
V_n\rightarrow\infty \iff |e_n|\rightarrow\infty,
\end{equation}
the error system \eqref{linearerrorA} is globally asymptotically stable whenever $|\rho|<1$. We emphasize that $|e_n|=|y_n-x_n|$ is constrained to approach infinity only via $y_n$ while the sequence $x_n$, remains in the basin of attraction. In other words, with the controller \eqref{controlA}, $y_n\in\mathbb{R}$ and $x_n\in(0,k)$. This is feasible as the GBS communication unit can be in a monitored environment, e.g. on a computer, while the ABS communication unit could be at a point in phase space where the physical dynamic range permits, yet, it is guaranteed to evolve toward stability.
\vspace{-5mm}
\section{UAV secure communication design}\label{five}
\begin{figure}[htbp]
	\centerline{\includegraphics[width=0.9\linewidth]{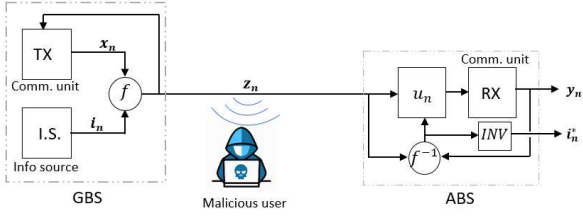}}
	\caption{Setup for information transmission in the UAS}
	\label{Fig3}
\end{figure}

The objective of the variable feedback controller in \cref{sync} is to drive the synchronization error to zero as the discrete-time sequence gets larger as given by \eqref{aim}. However, with the introduction of a sequence of information into the system, the above objective cannot be met. We exploit this supposedly negative outcome to create a useful product. \cref{Fig3} shows the setup for this process. Recall that states of the GBS and ABS communication units are $x_n$ and $y_n$, respectively. Generally, $x_n$ would evolve as:
\begin{equation}
x_n\rightarrow{x_n.o.i_n}=:z_n,
\end{equation}
where $i_n$ is a sequence of the information signal and $o$ is an invertible composition operator. Hence, we can define a new error as:
\begin{equation}
\epsilon_n := y_n - z_n.
\end{equation}
Provided our variable feedback controller continues to attempt to realize the originally intended objective in \eqref{aim}, then the following will be partially realized:
\begin{equation}
\lim\limits_{n\rightarrow\infty}\epsilon_n=i_n^*,
\end{equation}
where $i_n^*$ is a sequence with usually recognizable signatures of the original information signal. Therefore, a suitable detection mechanism can be used to reconstruct the original information signal. 

To further explain the setup in \cref{Fig3}, we note that the information signal $i_n$ is composed with $x_n$ of the GBS communication unit using any desired invertible function $f$, to get a scrambled signal $z_n$. The signal $z_n$ is being broadcast via a single channel and eavesdroppers might have access. However, they cannot recover the information in the signal as it is noise-like. Security of the information is based on the confidentiality of the logistic system, the variable feedback controller, including the scale factor and bounded feedback gain, and the unknown invertible function used. Besides, sensitive dependence on initial conditions plays a key role in the security. In some cases where the signature of the information in $i_n^*$ is relatively complicated, the detection mechanism could also play a role in the security. Besides, the synchronized units can be used to randomly select RF channels from a lookup table (see Appendix in \cref{randomRF}). 

At the receiving ABS unit, the variable feedback controller would use $z_n$ to synthesize a suitable control signal for the ABS communication unit. Outputs of the unit are compared with the incoming $z_n$ using the inverse function $f^{-1}$. The output of the function is the error $\epsilon$ which would give rise to $i_n^*$ after a few samples. At the stages where synchronization would be achieved assuming no information signal is input into the system, the signal $i_n^*$ would contain signatures of the original information signal. 
\vspace{-5mm}
\section{Simulation studies}\label{six}
\label{sim1}
\subsection{Information source turned off}
\label{sim1A}
\begin{figure}[htbp]
	\centerline{\includegraphics[width=0.9\linewidth]{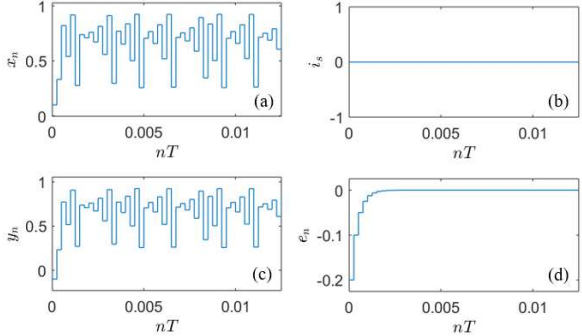}}
	\caption{\textbf{(a)} States of the transmitting unit. \textbf{(b)} Information source turned off. \textbf{(c)} States of the controlled receiver unit. \textbf{(d)} Synchronization error asymptotically converges to zero.}
	\label{Fig4}
\end{figure}

We assume the ABS and GBS are not exchanging data. They are only synchronized and ready to communicate. Hence, the information source as shown by \cref{Fig3} is turned off, i.e. $i_n=0$ for all $n$. Choosing parameter of the logistic map from within the chaotic space: $\mu = 3.7$ for example, using a scale factor, $k=1$, setting the initial condition of the drive system, $x_0 = 0.1$, and the response system's, $y_0 = -1.0$, setting the feedback gain, $\rho=0.5$, the sample time, $T = 2.5\times 10^{-4}$, and a runtime of $50T$, \cref{Fig4} shows results of the simulation with states of the ABS communication unit tracking those of the GBS communication unit as evidenced by asymptotic convergence to zero of the synchronization error. Notice that the ABS communication unit is initialized from outside the basin of attraction $(0,k)$. This indicates that the ABS communication unit tracks the GBS communication unit in such a manner that the error dynamical system is globally asymptotically stable. Besides, since the initial conditions are such that $x_0\neq y_0$ and $x_0+y_0\neq{k}$, the synchronization is due to the proposed feedback controller. Moreover, no information or data exchange is currently in session as indicated in \cref{Fig4}(b).
\subsection{Information source turned on}
\begin{figure}[htbp]
	\centerline{\includegraphics[width=0.9\linewidth]{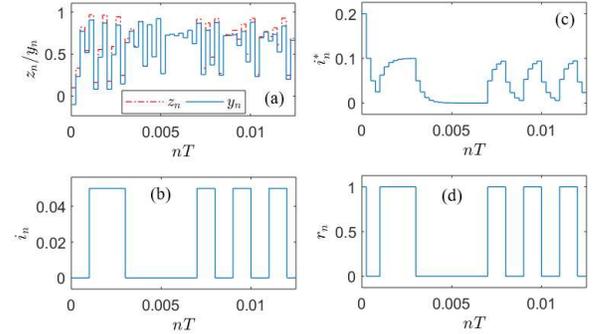}}
	\caption{\textbf{(a)} Transmitted scrambled signal $z_n$ and controlled output $y_n$, of the UAV communication unit. \textbf{(b)} The original information signal $i_n$. \textbf{(c)} Signal $i^*_n$ with signatures of $i_n$. \textbf{(d)} Recovered signal $r_n$.}
	\label{Fig5}
\end{figure}

In this case, the GBS is transmitting data to the ABS, i.e. $i_n\neq0$ for all $n$. Setting the communication units' parameters, including initial conditions, as in the previous case, using the Simulink function, Bernoulli Binary Generator as the information source, and using the invertible function of addition/subtraction, \cref{Fig5} shows the effectiveness of design. It can be seen in \cref{Fig5}(b) and (c) that $i_n^*$ has signatures of the information signal $i_n$. Using a simple threshold detection mechanism, \cref{Fig5}(d) shows the recovered signal $r_n$. During the first few samples before synchronization is achieved, it can be noticed that \cref{Fig5}(d) does not quite match with \cref{Fig5}(b). This can be avoided by applying a reasonable delay functionality but we have chosen to allow it because it is quite insightful.
\vspace{-5mm}
\section{FPGA prototype of the UAS communication units}\label{seven}
\subsection{Design approach}
\begin{figure}[htbp]
	\centering
	\includegraphics[width=0.9\linewidth]{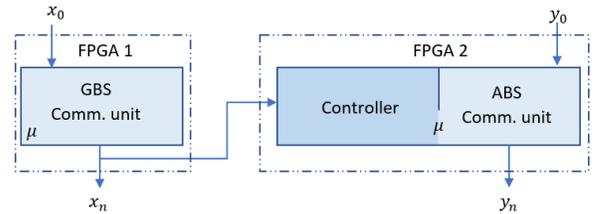}
	\caption{Prototyping the secure UAS communication.}
	\label{Fig6}
\end{figure}

As Fig.~\ref{Fig6} shows, states of the GBS communication unit can be used to synthesize a control signal based on the proposed variable feedback controller that synchronizes both units, irrespective of any difference in their initial conditions. Since $16$-bit resolution is sufficient to represent the dynamic range of the states, the $16$ switches of the Artix-7 FPGA board will suffice to pass in the initial conditions. We choose the scale factor $k = 2^{10}$. We are economical in the design approach due to our physical resource constraint. Hence, we coded the parameter $\mu$, into the design. In practice, it can be chosen randomly to heighten the security level.
The design is realized in the Simulink environment using function blocks that are compatible with hardware description language (HDL). Having satisfied fundamental digital design requirements such as sample rate and quantization, HDL codes are generated. 

\begin{figure}[htbp]
	\centerline{\includegraphics[width=0.9\linewidth]{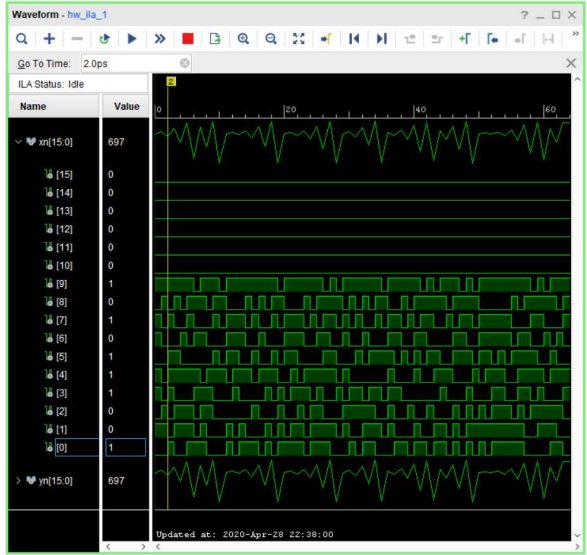}}
	\caption{Integrated logic analyzer showing the output signals from the Artix-7 FPGA. The current position of the cursor shows that the states of the drive and response systems are in sync.}
	\label{Fig7}
\end{figure}

The codes are further processed following the Xilinx Vivado design flow. From prior knowledge of the design slacks, using Tcl commands, we set the system clock to oscillate at $40~MHz$ (i.e. $25~ns$). This would ensure that timing constraints are met. Besides, the input-output delays are set appropriately so that before the completion of a clock cycle, the required input is in and the resultant output is out for reading. States of the drive and response system are routed to IO pins which will receive the digital signals. The configuration voltage is set to $1.8$ \textit{volts}. For ease of visualization, we created and connected a debug core to verify the logic states on the designated IO pins. 
Initial conditions for states of the drive and response system were `$0000000001111010$' (or $122$ in decimal) and `$1000010000000000$' (or $-1024$ in decimal), respectively. In
Fig.~\ref{Fig7} is a display captured on the Vivado Integrated Logic Analyzer. The current display is based on a data depth of $64$ samples. As the figure shows, the controller keeps the maps synchronized as the cursor position indicates a decimal value of $697$ (or `$0000001010111001$' in binary) for both outputs, $x_n$, and $y_n$. Allowing the FPGA to run for several hours, the implemented controller continues to ensure stable behavior. 
\vspace{-3mm}
\begin{remark}
	Although the error dynamical system is globally asymptotically stable, a confined dynamic range must be set due to hardware limitations.
\end{remark}
\subsection{Secure communication application}
We apply the synchronized logistic system to model secure exchanges of binary bitstreams between the GBS and ABS. The transmitter and receiver roles can be switched depending on whether a bitstream is outward-going or inward-going relative to the GBS (see Fig.~\ref{Fig1}). Hence, both the GBS and ABS have our proposed variable feedback controller. A transmitting device would broadcast only its states and the receiving device would use them to synthesize a control signal based on the proposed variable feedback controller. The control signal synchronizes the communication units when the information source is turned off. When it is turned on, although the variable feedback controller cannot attain absolute synchronization, the corresponding error would contain signatures of the information signal.

\begin{figure}[htbp]
	\centering
	{\includegraphics[width=0.9\linewidth]{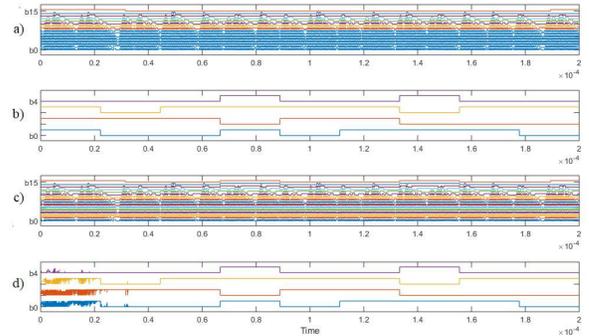}}
	\caption{Secure UAS communication application of the synchronized logistic system: a) 16-bit states of the transmitter. b) 4-bit information signal. c) 16-bit states of the unprocessed received signal. d) recovered signal.}
	\label{Fig8}
\end{figure}

To realize a digital simulation like in an FPGA, an $m$-bit sequence from the transmitting unit can be used to scramble an $n$-bit information signal, where $n\le{m}$. $n$ is required to be a factor of $m$, therefore, the quotient $r:=m/n$ is an integer. For each sample depth, $r$ bits of the transmitter state is used to scramble a bit of the information signal. Assuming the communication channel is ideal, the synchronized response system or receiver can be used to descramble the signal. The received signal $R_s=\{1,1,1,1,0,0,0,0,...1\}$ for example, can be $m$-bit long. Hence, $R_s$ would be passed through a correlator with the following operation:
\begin{equation}
s_p={1\over r}\sum_{k={r}(p-1)+1}^{{r}p}R_s(k) = 
\begin{cases}
1 & \forall\enspace R_s(k) = 1\\
0 & \forall\enspace R_s(k) = 0\\
(0,1)&\quad otherwise\label{digcorrelation},
\end{cases}
\end{equation}
for each $p=1,2,..,n$, and $\forall$ strictly means $\mathit{for~all}$. Equation \eqref{digcorrelation} performs a sequential collection of $r$ elements from the received signal and averages over $r$. The descrambled signal will have $n$-bit resolution and comprise a sequential collection of $s_p$. Fig.~\ref{Fig8} shows effectiveness of the proposed method. Earlier on in Fig.~\ref{Fig8}(d), before synchronization is reached, fringes can be seen in the recovered signal. Afterwards, the recovered signal identifies the original information signal, Fig.~\ref{Fig8}(b), as time progresses.
\vspace{-3mm}
\vspace{-5mm}
\section{Conclusion and summary}
This work proposes a secure communication architecture that uses deterministic chaos to realize secure UAV communication. The circuitry of the communicating units viz., the ground base station (GBS) and a UAV identified as the aerial base station (ABS), is based on the logistic map in the chaotic domain. To synchronize the units, we propose a variable feedback controller and prove global asymptotic stability of the error dynamics given that no transmission is in session. However, when the GBS for example, transmits information, there is only stability but not asymptotic stability. We exploit this seemingly suboptimal outcome to realize useful secure communication between the GBS and ABS. Besides, we realize an FPGA prototype of the synchronized communication units. We also realize a digital design simulation targeting UAS security that involves the transmission of bitstreams from the GBS to the ABS. The success of the method is based on a simple correlation summation algorithm. The various methods are verified in numerical simulations and FPGA prototyping. Future research would seek to realize a robust controller to synchronize the logistic system for secure UAV-to-UAV communication via a channel with modeled disturbances.
\vspace{-5mm}
\appendix{}  
\subsection{Random RF channel selection}
\label{randomRF}
\begin{figure}[htbp]
	\centering
	\includegraphics[width=\linewidth]{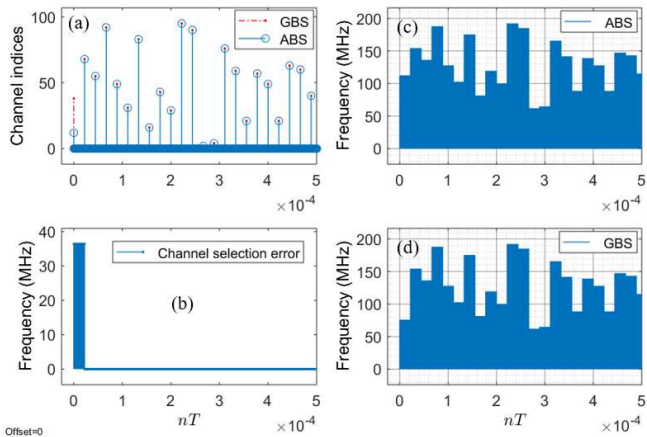}
	\caption{Sychronized channel selection following chaos-based pseudo-random sequence}
	\label{Fig9}
\end{figure}

We show that due to synchronization of the GBS and ABS by the proposed variable feedback controller, it is possible to realize automatic random-like RF channel selection. The selection can alternate between sessions of information transmission states of \textit{on} and \textit{off}. When information transmission is \textit{off}, there is absolute synchronization of the GBS and ABS as seen in \cref{sim1A}. During this period, a channel selection can be triggered to move away from the previous channel. It should be noted that the selection follows no identifiable pattern as it is completely based on states of the logistic map in the chaotic domain. As a channel index $j$, is selected from the onset of synchronization as shown in \cref{Fig9}(a), the GBS and ABS will communicate on a specific RF channel determined from a lookup table (LUT) as can be seen in \cref{freqhopping} and \cref{Fig9}(c) and (d). The channel selection error remains zero after synchronization is realized for the GBS and ABS communication units as \cref{Fig9}(b) shows.  

Although channel indices in the table are sequential, their selection is not. The selection is based on states of the logistic map in the chaotic domain for which there is the impossibility of long-term prediction. This is yet another security layer of our proposed method. Therefore, suppose an eavesdropper happens to stumble on the current GBS and ABS communication channel, in the next session of information exchange, only white noise would be received by the eavesdropper as communication would be established on a different randomly selected channel.
\vspace{-1mm}
\begin{table}[htbp]
	\caption{\bf Frequency lookup table (LUT)}
	\label{freqhopping}
	\centering
	\begin{tabular}{|c|c|c|}
		\hline
		\bfseries Channel & \bfseries Frequency & \bfseries Center\\
		\bfseries indices & \bfseries range & \bfseries frequency\\
		\bfseries  $j$ & \bfseries ($MHz$) & \bfseries ($MHz$)\\
		\hline\hline
		1&$60.0-61.4$&$60.7$\\
		2&$61.4-62.8$&$62.1$\\
		.&.&.\\
		.&.&.\\
		99&$197.2-198.6$&$197.9$\\
		100&$198.6-200.0$&$199.3$\\
		\hline
	\end{tabular}
\end{table}
\vspace{-5mm}
\section*{Acknowledgment}
The authors appreciate the support of the College of Engineering, North Carolina State University, USA. The first author appreciates the useful discussions with Dr. Mario Aldape P\'erez at CIDETEC of National Polytechnic Institute, Mexico.

\bibliography{ieeeAerospaceconf}

\begin{thebibliography}{10}
\providecommand{\url}[1]{#1}
\csname url@samestyle\endcsname
\providecommand{\newblock}{\relax}
\providecommand{\bibinfo}[2]{#2}
\providecommand{\BIBentrySTDinterwordspacing}{\spaceskip=0pt\relax}
\providecommand{\BIBentryALTinterwordstretchfactor}{4}
\providecommand{\BIBentryALTinterwordspacing}{\spaceskip=\fontdimen2\font plus
\BIBentryALTinterwordstretchfactor\fontdimen3\font minus
  \fontdimen4\font\relax}
\providecommand{\BIBforeignlanguage}[2]{{%
\expandafter\ifx\csname l@#1\endcsname\relax
\typeout{** WARNING: IEEEtran.bst: No hyphenation pattern has been}%
\typeout{** loaded for the language `#1'. Using the pattern for}%
\typeout{** the default language instead.}%
\else
\language=\csname l@#1\endcsname
\fi
#2}}
\providecommand{\BIBdecl}{\relax}
\BIBdecl

\bibitem{shakhatreh2019unmanned}
H.~{Shakhatreh}, A.~H. {Sawalmeh}, A.~{Al-Fuqaha}, Z.~{Dou}, E.~{Almaita},
  I.~{Khalil}, N.~S. {Othman}, A.~{Khreishah}, and M.~{Guizani}, ``Unmanned
  aerial vehicles (uavs): A survey on civil applications and key research
  challenges,'' \emph{IEEE Access}, vol.~7, pp. 48\,572--48\,634, Apr. 2019.

\bibitem{yaacoub2020security}
J.-P. Yaacoub and O.~Salman, ``Security analysis of drones systems: Attacks,
  limitations, and recommendations,'' \emph{Internet of Things}, vol.~11, pp.
  1--39, May 2020.

\bibitem{ezuma2019micro}
M.~Ezuma, F.~Erden, C.~K. Anjinappa, O.~Ozdemir, and I.~Guvenc, ``Micro-{UAV}
  detection and classification from rf fingerprints using machine learning
  techniques,'' in \emph{Proc. IEEE Aerosp. Conf., Big Sky, Montana}, Mar.
  2019, pp. 1--13.

\bibitem{ezuma2019detection}
------, ``Detection and classification of {UAVs} using {RF} fingerprints in the
  presence of {Wi-Fi} and {B}luetooth interference,'' \emph{IEEE OJ-COMS},
  vol.~1, pp. 60--76, Nov. 2019.

\bibitem{IDISOUAV2019}
\BIBentryALTinterwordspacing
ISO, ``(21384-3:2019) unmanned aircraft systems -- part 3: Operational
  procedures,'' Nov. 2019. [Online]. Available:
  \url{https://www.iso.org/standard/70853.html}
\BIBentrySTDinterwordspacing

\bibitem{faraji2020secure}
M.~Faraji-Biregani and R.~Fotohi, ``Secure communication between uavs using a
  method based on smart agents in unmanned aerial vehicles,'' \emph{J.
  Supercomput.}, pp. 1--28, Nov. 2020.

\bibitem{pecora2015synchronization}
L.~M. Pecora and T.~L. Carroll, ``Synchronization of chaotic systems,''
  \emph{Chaos}, vol.~25, no.~9, pp. 1--12, Apr. 2015.

\bibitem{bendoukha2019secure}
S.~Bendoukha, S.~Abdelmalek, and A.~Ouannas, ``Secure communication systems
  based on the synchronization of chaotic systems,'' in \emph{Mathematics
  Applied to Engineering, Modelling, and Social Issues}.\hskip 1em plus 0.5em
  minus 0.4em\relax Springer, Mar. 2019, pp. 281--311.

\bibitem{yu2019survey}
F.~Yu, L.~Li, Q.~Tang, S.~Cai, Y.~Song, and Q.~Xu, ``A survey on true random
  number generators based on chaos,'' \emph{Discrete Dyn. Nat. Soc.}, vol.
  2019, pp. 1--10, Dec. 2019.

\bibitem{moysis2020chaotic}
L.~Moysis, E.~Petavratzis, C.~Volos, H.~Nistazakis, and I.~Stouboulos, ``A
  chaotic path planning generator based on logistic map and modulo tactics,''
  \emph{Rob. Auton. Syst.}, vol. 124, pp. 1--25, Feb. 2020.

\bibitem{may1976simple}
R.~M. May, ``Simple mathematical models with very complicated dynamics,''
  \emph{Nature}, vol. 261, no. 5560, pp. 459--467, Jun. 1976.

\bibitem{lorenz1963deterministic}
E.~N. Lorenz, ``Deterministic nonperiodic flow,'' \emph{J. Atmos. Sci.},
  vol.~20, no.~2, pp. 130--141, Mar. 1963.

\bibitem{holyst1996control}
J.~A. Holyst, T.~Hagel, G.~Haag, and W.~Weidlich, ``How to control a chaotic
  economy?'' \emph{J. Evol. Econ.}, vol.~6, no.~1, pp. 31--42, Mar. 1996.

\bibitem{Nwachioma2019New}
C.~{Nwachioma}, J.~{Humberto Pérez-Cruz}, A.~{Jiménez}, M.~{Ezuma}, and
  R.~{Rivera-Blas}, ``A new chaotic oscillator—properties, analog
  implementation, and secure communication application,'' \emph{IEEE Access},
  vol.~7, pp. 7510--7521, Jan. 2019.

\bibitem{nwachioma2017realization}
C.~Nwachioma and J.~H. P{\'e}rez-Cruz, ``Realization and implementation of
  polynomial chaotic sun system,'' \emph{Phys. Sci. Int. J.}, vol.~16, no.~4,
  pp. 1--7, Jan. 2017.

\bibitem{can2020new}
E.~Can, U.~E. Kocamaz, and Y.~Uyaro{\u{g}}lu, ``A new six-term 3d unified
  chaotic system,'' \emph{IJST-T Electr. Eng.}, vol.~44, no.~4, p. 1593–1604,
  Feb. 2020.

\bibitem{sprott2010elegant}
\BIBentryALTinterwordspacing
J.~C. Sprott, \emph{Elegant Chaos: Algebraically Simple Chaotic Flows}.\hskip
  1em plus 0.5em minus 0.4em\relax World Sci. Pub. Co. Plc Ltd., Mar. 2010,
  ch.~1, pp. 10--29. [Online]. Available:
  \url{https://www.worldscientific.com/doi/abs/10.1142/7183}
\BIBentrySTDinterwordspacing

\bibitem{ali1997ssynchronization}
M.~Ali, ``Synchronization of a chaotic map in the presence of common noise,''
  \emph{Phys. Rev. E}, vol.~55, no.~4, pp. 4804--4805, Apr. 1997.

\bibitem{morgul1998synchronization}
{\"O}.~Morg{\"u}l, ``On the synchronization of logistic maps,'' \emph{Phys.
  Lett. A}, vol. 247, no.~6, pp. 391--396, Oct. 1998.

\bibitem{perez2020luenberger}
J.~H. P{\'e}rez-Cruz, J.~M. Allende~Pe{\~n}a, C.~Nwachioma, J.~d.~J. Rubio,
  J.~Pacheco, J.~A. Meda-Campa{\~n}a, D.~{\'A}vila-Gonz{\'a}lez,
  O.~Guevara~Galindo, I.~Adrian~Romero, and S.~I. Belmonte~Jim{\'e}nez, ``A
  luenberger-like observer for multistable kapitaniak chaotic system,''
  \emph{Complexity}, vol. 2020, pp. 1--12, Jul. 2020.

\end{thebibliography}
\bibliographystyle{IEEEtran}


\thebiography
\begin{biographywithpic}
	{Christian Nwachioma}{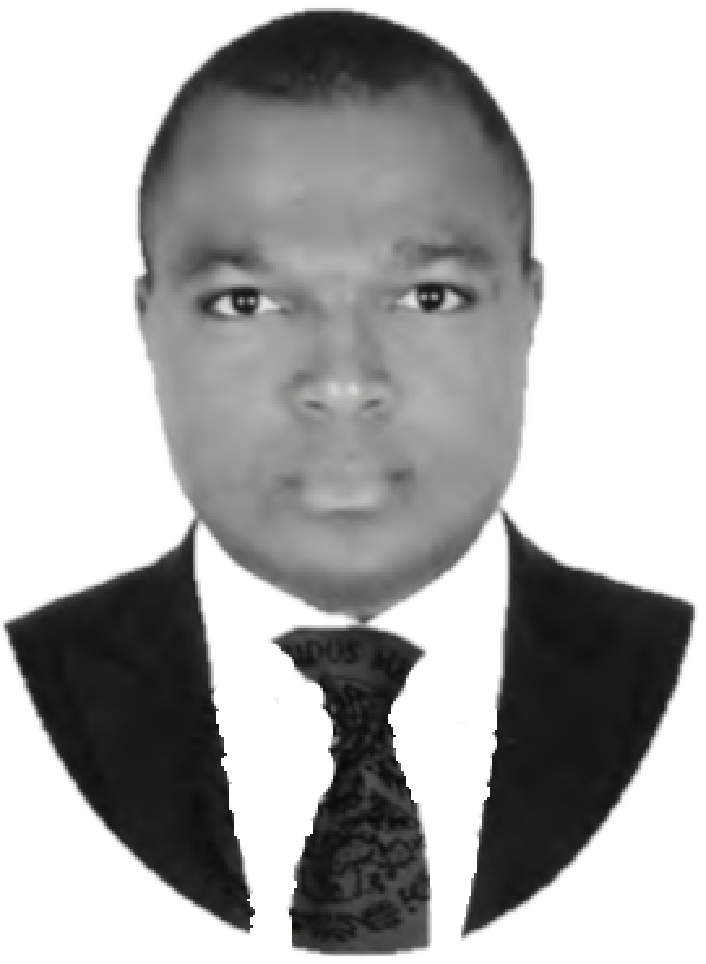}
	received his B.S. and M.S. degrees in Physics from Federal University of Technology, Owerri, Nigeria, in 2010 and COMSATS University, Islamabad, Pakistan, in 2016, respectively. He recently received a doctorate in Robotics and Mechatronic System Engineering from Instituto Polit\'ecnico Nacional, Mexico, in July 2020. His current research interest includes Control theory, robotics and control, and applied computation.
\end{biographywithpic} 
\begin{biographywithpic}
	{Martins Ezuma}{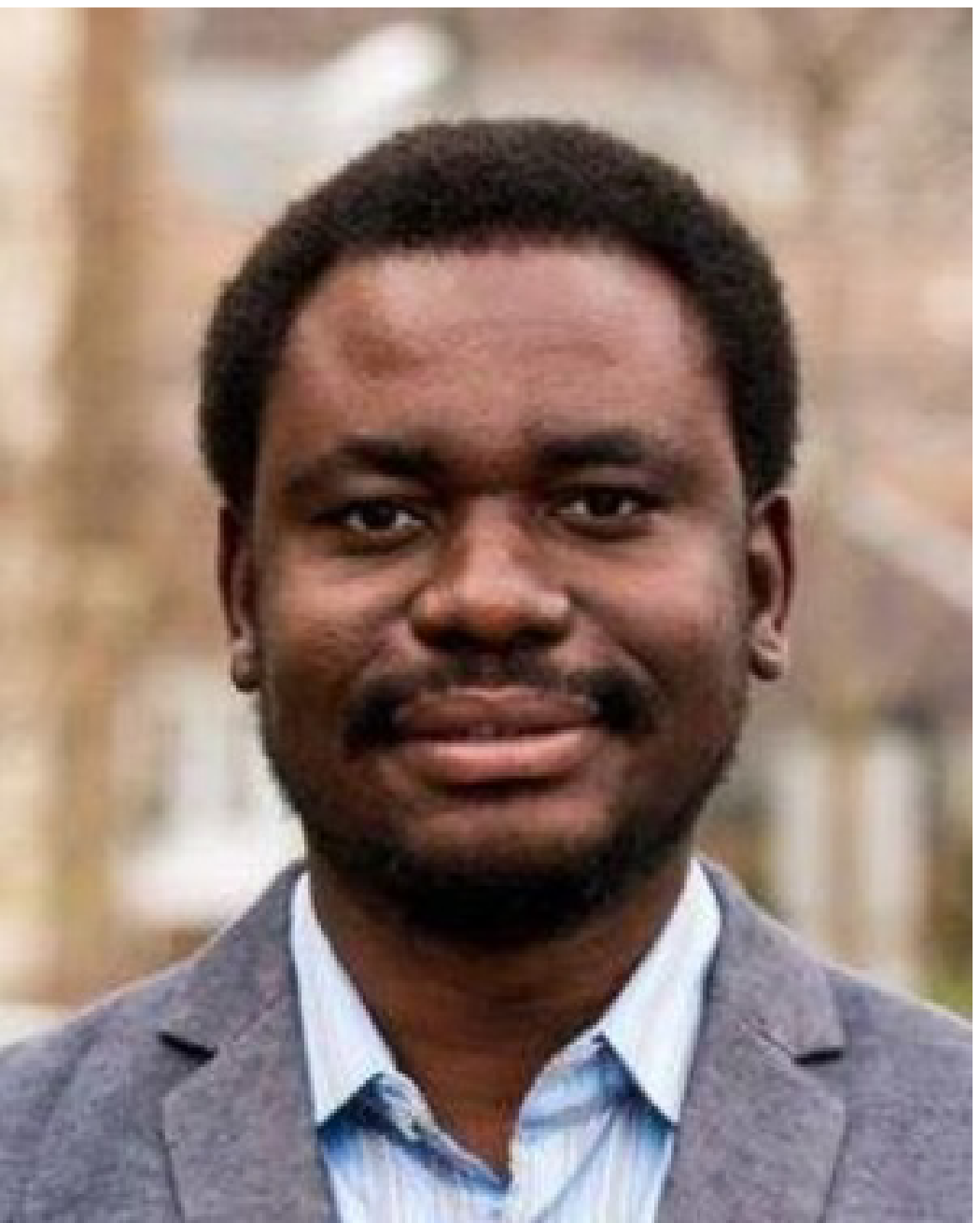}
	received BS in Physics from Federal University of Technology, Owerri, Nigeria in 2010, MS in Information and Communication Engineering from Chosun University, Gwangju, South Korea in 2015, and an MS degree in electrical engineering from New Jersey Institute of Technology (NJIT), in 2016. His research interest include signal processing, 5G channel sounding and pattern recognition. He is currently pursuing his Ph.D. degree in Electrical and Computer Engineering at North Carolina State University (NCSU).
\end{biographywithpic}
\begin{biographywithpic}
	{Olusiji .O. Medaiyese}{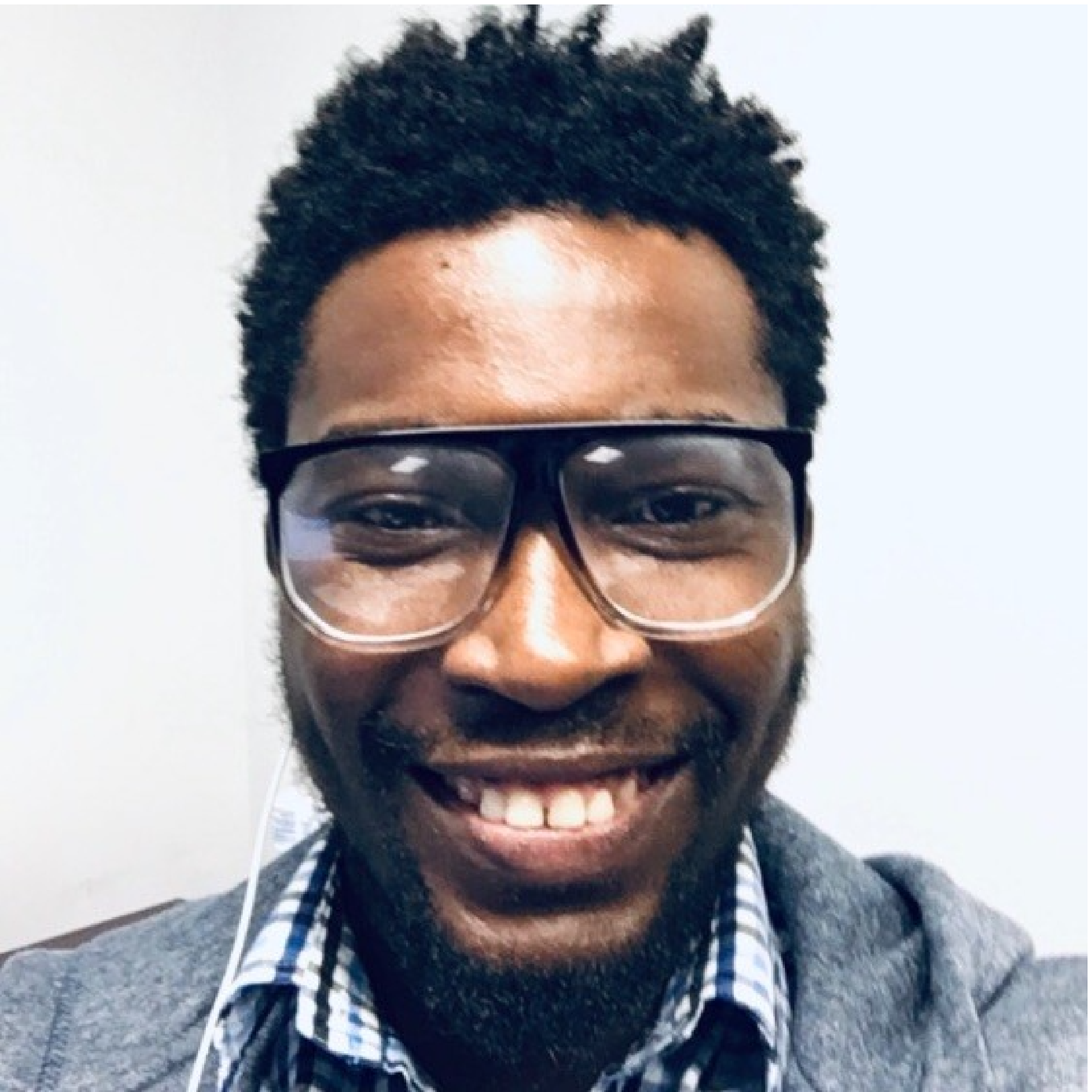}
	received B.Eng. in Electrical and Electronics Engineering from the University of Ilorin, Ilorin, Nigeria in 2012. MS in Computer and Systems Engineering from Tallinn University of Technology, Tallinn, Estonia in 2018. He is currently pursuing his Ph.D. degree in Computer Science and Engineering at the University of Louisville, Louisville, Kentucky, USA. His research area is machine learning with signal processing, mobile Adhoc network, and UAV detection and identification.
\end{biographywithpic}
\end{document}